\documentclass[twocolumn,pra,superscriptaddress,noeprint]{revtex4}%
\usepackage{amsfonts}
\usepackage{amsmath}
\usepackage{amssymb}
\usepackage{graphicx}%
\begin{document}
\title{Enhancing test precision for local Lorentz symmetry violation with entanglement}
\author{Lei Li}
\affiliation{School of Mathematics and Physics, China University of Geosciences, Wuhan
430074, China}
\author{Xinwei Li}
\affiliation{State Key Laboratory of Low Dimensional Quantum Physics, Department of
Physics, Tsinghua University, Beijing 100084, China}
\author{Baocheng Zhang}
\email{zhangbc.zhang@yahoo.com}
\affiliation{School of Mathematics and Physics, China University of Geosciences, Wuhan
430074, China}
\author{Li You}
\affiliation{State Key Laboratory of Low Dimensional Quantum Physics, Department of
Physics, Tsinghua University, Beijing 100084, China}
\keywords{Lorentz symmetry violation, local interaction, Einstein equivalence principle,
classical violation}
\pacs{11.30.-j, 04.20.-q, 04.80.Cc }

\begin{abstract}
A recent proposal for testing Lorentz symmetry violation (LSV) presents a
formulation where the effect of violation is described as a local interaction
[R. Shaniv, et al, Phys. Rev. Lett. \textbf{120}, 103202 (2018)]. An entangled
ion pair in a decoherence free subspace (DFS) is shown to double the signal to
noise ratio (SNR) of one ion, while (even)-N/2 such DFS pairs in a collective
entangled state improve SNR by N times, provided the state parity or the
even/odd numbers of ions can be measured. It remains to find out, however, how
such fiducial entangled states can be prepared at nonexponentially small
success rates. This work suggests two types of many particle entangled states
for testing LSV: the maximally entangled NOON state, which can achieve
Heisenberg limited precision; and the balanced spin-1 Dicke state, which is
readily available in deterministic fashion. We show that the latter also lives
in a DFS and is immune to stray magnetic fields. It can achieve classical
precision limit or the standard quantum limit (SQL) based on collective
population measurement without individual atom resolution. Given the high
interests in LSV and in entanglement assisted quantum metrology, our
observation offers additional incentives for pursuing practical applications
of many atom entangled states.

\end{abstract}
\maketitle

\section{Introduction}

Invariance under Lorentz transformation constitutes one of the most
fundamental principles of modern physics. A more complete theory including
quantum gravity, however, implies the possibility or even necessity for
Lorentz symmetry violation (LSV) \cite{dm05,sl13,jdt14}. Hence, the study of
this violation has attracted wide attention. Over the past twenty years a
theoretical framework called the Standard-Model Extension (SME)
\cite{cm97,cm98,kr11} is developed, which formally contains all possible
symmetry breaking terms generated by couplings between the standard fields and
the vacuum expectations of the tensor fields parameterizing symmetry
violations. Within the SME, the symmetry breaking terms collectively do not
cause LSV in the Planck scale. But in low energy scale (due to spontaneously
symmetry breaking) every term causes violation in the presumed absence of the
others \cite{cm98}, and can be studied experimentally.

Of direct relevance to this study concerns several recent efforts using atoms
or ions to test LSV through precision measurements. The specific LSV term
usually concerns the isotropy of the speed of light (for recent results,
please refer to Ref. \cite{hsm09}), with the most sensitive tests to date
using neutral Dy atoms \cite{hlb13}, Ca$^{+}$ ions \cite{prp15}, and Yb$^{+}$
ions \cite{dfh16}. An alternative proposal recently suggests testing atomic
level spacing with atomic clock \cite{sos18} (augmented by dynamic
decoupling), predicting an extremely high sensitivity to LSV. Yet another term
in the SME concerns the dependence on the direction of motion and the momentum
of a particle, which also violates local Lorentz invariance (LLI)
\cite{mtw70}. If the total angular momentum of a physical system is fixed,
this latter term becomes proportional to the square of the $z$-component of
the total angular momentum operator, as in quadratic Zeeman shift for an
atomic hyperfine spin, which constitutes a typical local interaction
\cite{sos18}.

The first two types of tests mentioned above all reduce to atomic clock
measurements. Their precisions can reach beyond the classical precision limit,
or the standard quantum limit (SQL) $(\propto1/\sqrt{N}$), when ensembles of
entangled particles are employed \cite{wineland}. Our studies indicate that
the third type of term, an equivalent quadratic Zeeman shift or a local
interaction form, can also be tested beyond the SQL with entangled particles.

This paper is organized as follows. First we introduce the relevant background
as well as the formulation for the specific LSV term (from the SME) we study.
This is followed by a discussion of the recent experimental proposal for
estimating its strength with the help of a DFS. Our main contribution is then
presented by suggesting two types of many atom entangled states instead.
First, the maximally entangled NOON state, with which Heisenberg limited (HL)
precision $\propto1/N$ becomes possible; Second, the balanced spin-1 Dicke
state \cite{zwl18}, which lives in its own DFS and is thus also immune to
stray magnetic fields, can be employed to reach the SQL based on collective
population measurement without requiring individual atom resolution.

\section{Lorentz symmetry violation}

Based on the modern description of nature at the most fundamental level,
Lorentz symmetry might be violated at experimentally accessible energy scales
due\ to spontaneous symmetry breaking, hence the development of its
experimental test \cite{kl99,kr99,mhs04,hsm09,sos18,hlb13,kv15,prp15,dfh16}.
In relativistic physics, Lorentz symmetry implies an equivalence of
observation or observational symmetry according to special relativity, which
is formally equivalent to stating that the laws of physics stay the same for
all observers moving at constant velocities with respect to each other in
inertial frames. It is also described as the independence of experimental
results on orientation or boost velocity of the apparatus setup through space
\cite{mtw70}. When LSV is considered for the electron sector within the
framework of SME, the QED Lagrangian (for electron) becomes \cite{cm97,cm98},
\begin{equation}
L=\frac{1}{2}i\,\overline{\psi}\left(  \gamma_{\nu}+c_{\mu\nu}\gamma^{\mu
}\right)  \overleftrightarrow{D}^{\nu}\psi-\overline{\psi}\,m_{e}\psi,
\label{lag}%
\end{equation}
where $m_{e}$ denotes the electron mass, $\psi$ is a Dirac spinor,
$\gamma^{\mu}$ are the usual Dirac matrices and $\overline{\psi}%
\overleftrightarrow{D}^{\nu}\psi\equiv\overline{\psi}D^{\nu}\psi-\psi
D^{\upsilon}\overline{\psi}$ with $D^{\upsilon}$ the shorthand for the
covariant derivative. The $c_{\mu\nu}$ tensor in the above Eq. (\ref{lag})
quantifies the strength of LSV for the electron sector by the frame dependent
interaction term, which gives an energy level shift \cite{kl99,kl992,kt11}:
$\delta H_{\mathrm{LSV}}=-[C_{0}^{(0)}-{2U}c_{00}/({3c^{2})}]{\mathbf{p}^{2}%
}/{2}-C_{0}^{(2)}T_{0}^{(2)}/6$, responsible for the first two types of LSV
tests mentioned in the last section. Here, $U$ is the Newtonian gravitational
potential and the more specific parameters $C_{0}^{(0)}$ and $c_{00}$
quantifying the strength of LSV have been discussed and tested before
\cite{hlb13}, but are not required for the following discussions.

The relativistic form of rank 2 irreducible tensor operator $T_{0}^{(2)}$ is
$T_{0}^{(2)}=c\gamma_{0}(\mathbf{\gamma p}-3\gamma_{z}p_{z})$, with $p_{z}$
the momentum component along the quantization axis fixed in the laboratory
frame. Its non-relativistic form becomes $T_{0}^{(2)}=({\mathbf{p}^{2}%
-3p_{z}^{2}})/{m_{e}}$. Thus the second LSV term $\propto C_{0}^{(2)}$ of
$\delta H_{\mathrm{LSV}}$ reduces to
\begin{equation}
\delta H=-C_{0}^{(2)}\frac{\left(  \mathbf{p}^{2}-3p_{z}^{2}\right)  }{6m_{e}%
},\label{eht}%
\end{equation}
which is responsible for breaking of the symmetry on independence of
orientation or velocity in a bound electron system such as a Ca$^{+}$ ion
\cite{prp15}. Its diagonal matrix element
\begin{equation}
\langle j,m|T_{0}^{(2)}|j,m\rangle=\frac{[-j(j+1)+3m^{2}]\langle
j|T^{(2)}|j\rangle}{\sqrt{(2j+3)(j+1)(2j+1)j(2j-1)}},\label{wet}%
\end{equation}
is calculated using Wigner-Eckart theorem \cite{prp15}, with $j$ and $m$ the
corresponding quantum numbers for the total and z-component electron angular
momentum. The latter ($\propto m^{2}$ term) corresponds to raw LSV signal. For
a physical system with a fixed $j$, such as a bound electron system first
suggested in Ref. \cite{sos18}, the LSV dynamics is thus described by an
equivalent Hamiltonian $H_{V}=\kappa j_{z}^{2}$, which is analogous to a
quadratic Zeeman shift.

In earlier LSV tests of the local interaction term Eq. (\ref{eht}),
eigenstates of distinct absolute angular momentum $j_{z}$ ($=m$) are chosen in
order to extract the relative (time) phase from a coherent superposition, over
days or even longer times. In Ref. \cite{sos18}, the selected states are
$\left\vert \frac{7}{2},-\frac{7}{2}\right\rangle $ and $\left\vert \frac
{7}{2},-\frac{1}{2}\right\rangle $, their relative time phase is measured by
Ramsey interferometry, sometimes, augmented by dynamical decoupling (DD)
\cite{lb13}, as described in the following. In terms of the total angular
momentum operator for a particle, the time evolution operator for the Ramsey
interferometry becomes \cite{nr85,ymk86}
\[
U_{\phi}=e^{-i\pi j_{x}/2}e^{i\phi j_{z}}e^{i\pi j_{x}/2}=e^{-i\phi j_{y}},
\]
with the free evolution term $e^{i\phi J_{z}}$ (rotation of an angle $\phi$
around the $z$-axis) sandwiched in between two ${\pi}/{2}$ pulses that
effectively serve as 50:50 beam splitters. Thus the relative phase $\phi$ maps
onto the familiar differential optical path of the Mach-Zehnder
interferometry. With the LSV term included into free evolution, the total time
evolution changes into
\[
U_{\phi,\kappa}=e^{-i\pi j_{x}/2}e^{i\phi j_{z}}e^{-i\kappa tj_{z}^{2}}e^{i\pi
j_{x}/2}=e^{-i\phi j_{y}-i\kappa tj_{y}^{2}},
\]
where the second term in the exponent above $\propto\kappa$ resembles the
one-axis spin squeezing \cite{ku93} or its associated quadratic interaction in
nonlinear Ramsey interferometry \cite{uf03,cb05,cs08}. A careful examination
reveals, however, that this is simply not the case as the quadratic spin
operator $j_{z}^{2}$ here refers to one particle rather than the collective
spin of an ensemble \cite{ku93}.

It is well known according to parameter estimation theory \cite{cwh76,mgp09}
that an ensemble of uncorrelated particles, even in the ideal case, provides
the best achievable precision for estimating $\delta\kappa$ $\propto1/\sqrt
{N}$, consistent with the SQL. The total LSV Hamiltonian for an ensemble of
$N$ atoms can be expressed as
\begin{align}
H_{V}=\kappa(j_{z}^{(1)})^{2}+\kappa(j_{z}^{(2)})^{2}\cdots+\kappa(j_{z}%
^{(N)})^{2}\equiv\kappa\mathcal{H},\hskip12pt \label{Hes}%
\end{align}
where $j_{z}^{(i)}$ denotes $j_{z}$ for the i-th atom, and $\mathcal{H}%
=\sum\limits_{i=1}^{N}(j_{z}^{(i)})^{2}$ is the generator for estimating
$\kappa$. Evidently, the ensemble evolution operator is
\begin{align}
U_{\phi,\kappa}=\prod_{i=1}^{N}e^{-i\phi j_{y}^{(i)}-i\kappa t(j_{y}%
^{(i)})^{2}},
\end{align}
which remains local as $H_{V}$ induces no entanglement if all atoms are
initially in product states. Nevertheless, various techniques can be
implemented for improved estimation of $\kappa$ based on simple population
measurements in spin components in the end \cite{sos18}.

Quantum estimation theory \cite{cwh76,mgp09} allows for the precision of
estimating a local parameter to go beyond the SQL, \textit{e.g.} with quantum
entangled ensembles. For the LSV term \cite{sos18} we discuss, it was pointed
out that an entangled pure state of two Yb$^{+}$ ions,
\begin{align}
{\frac{1}{\sqrt{2}}}\left(  \left\vert {\frac{7}{2}},{\frac{7}{2}%
}\right\rangle \left\vert {\frac{7}{2}},-{\frac{7}{2}}\right\rangle
+\left\vert {\frac{7}{2}},{\frac{1}{2}}\right\rangle \left\vert {\frac{7}{2}%
},-{\frac{1}{2}}\right\rangle \right)  , \label{dfspair}%
\end{align}
gives a factor of $\sqrt{2}$ improvement in the signal to noise ratio (SNR),
beating the SQL \cite{prp15}. In addition, such a two-ion state forms a
decoherence free subspace (DFS), which significantly suppresses stray magnetic
fields. Extrapolating further (in the supplemental materials of Ref.
\cite{prp15}), an analogous paired entangled state
\begin{equation}
{\frac{1}{\sqrt{2}}}\left(  \left\vert {\frac{7}{2}},{\frac{7}{2}%
}\right\rangle ^{\otimes{\frac{N}{2}}}\left\vert {\frac{7}{2}},-{\frac{7}{2}%
}\right\rangle ^{\otimes{\frac{N}{2}}}+\left\vert {\frac{7}{2}},{\frac{1}{2}%
}\right\rangle ^{\otimes{\frac{N}{2}}}\left\vert {\frac{7}{2}},-{\frac{1}{2}%
}\right\rangle ^{\otimes{\frac{N}{2}}}\right)  , \label{Npairs}%
\end{equation}
is shown to provide $N$-times enhanced precision (for $N$ even). The two
states in Eqs. (\ref{dfspair}) and (\ref{Npairs}) provide enhanced estimation
of $\kappa$, but require parity measurements, i.e., the ability to determine
even/odd of the total numbers of ions projected into a specific internal
state, although Ramsey interferometries are applied to all ions at once. The
required parity measurement, however, represents a serious constrain against
the most significant advantages for ensemble based precision measurement:
collective manipulations and measurements, with Ramsey pulses applied to all
particles at once indiscriminately and total populations in different single
particle internal states counted in the end without single atom resolution. In
the following, we suggest two alternative multi-particle entangled states for
estimating $\kappa$ with enhanced precision.

\section{NOON state and the balanced spin-1 Dicke state}

The state Eq. (\ref{Npairs}) is very suggestive, as it represents nothing but
an equal superposition of two twin-Fock states \cite{luo2017deterministic}.
Each of the twin-Fock modes satisfies $m=-m$, offering the combined advantages
of: DFS or immunity (or insensitivity) to stray magnetic fields, and a
cumulative relative $m^{2}$-dependent LSV phase. For any meaningfully large
$N$, however, it is unclear how such a state can be generated with high yield
as the method of Ref. \cite{prp15,dfh16,sos18} is exponentially ineffective.

More generally, estimating $\kappa$ starts with encoding an input state $\rho$
by a unitary transformation $U = {e^{ - i\kappa t\mathcal{H}}}$ and follows
with measurements to the output state ${\rho_{\kappa}} = U\rho U^{\dag}$. The
ultimate precision is given by the quantum Cram\'{e}r-Rao bound (QCRB)
\cite{bc94,bc96}
\begin{align}
\delta\kappa\ge{\frac{1 }{\sqrt\nu}}{\frac{1 }{{T\sqrt{{F_{Q}}} }}},
\label{QFI}%
\end{align}
where ${{F_{Q}}}$ denotes quantum Fisher information (QFI) and is determined
by the output state ${\rho_{\kappa}}$, $T$ is the time duration of an ensemble
state (used to probe) under $H_{V}$ and $\nu$ is the number of experimental
trials. If the probe state is pure $\left|  \psi\right\rangle $, QFI
simplifies to
\begin{align}
{F_{Q}} = 4{(\Delta\mathcal{H})^{2}} = 4( {\left\langle {{\mathcal{H}^{2}}}
\right\rangle - {{\left\langle \mathcal{H} \right\rangle }^{2}}}),
\label{QFIpure}%
\end{align}
which gives the QCRB for a pure probe state \cite{bdf08}
\begin{align}
\delta\kappa\geq{\frac{1}{\sqrt{\nu}}}{\frac{1}{{2T\Delta\mathcal{H}}}},
\label{QCRB}%
\end{align}
with the variance $(\Delta\mathcal{H})^{2}={\left\langle {{\mathcal{H}^{2}}%
}\right\rangle -{{\left\langle \mathcal{H}\right\rangle }^{2}}}$, bounded by
\begin{align}
\Delta\mathcal{H}\leq{{({{\Lambda_{\max}}-{\Lambda_{\min}}})}/{2}}.
\label{del_H}%
\end{align}
${{\Lambda_{\max}}}$ and ${{\Lambda_{\min}}}$ are the maximum and minimum
eigenvalues of $\mathcal{H}$ \cite{bc94}. This bound can be achieved by using
the initial probe state $\left(  {\left\vert {{\Lambda_{\max}}}\right\rangle
+\left\vert {{\Lambda_{\min}}}\right\rangle }\right)  /\sqrt{2}$, a cat state
with $\left\vert {{\Lambda_{\max}}}\right\rangle $ ($\left\vert {{\Lambda
_{\min}}}\right\rangle $) the eigenstate of corresponding eigenvalue
${{\Lambda_{\max}}}$ (${{\Lambda_{\min}}}$). The maximum and minimum
eigenvalue for $\mathcal{H}$ defined in Eq. (\ref{Hes}) is ${N{\lambda_{\max}%
}}$ and ${N{\lambda_{\min}}}$ respectively, where ${{\lambda_{\max}}}$
(${{\lambda_{\min}}}$) denotes the maximum (minimum) eigenvalue of the single
particle $(j_{z}^{(i)})^{2}$. Thus the QCRB is given by
\begin{align}
\delta\kappa\geq{\frac{1}{\sqrt{\nu}}}{\frac{1}{{TN\left\vert {\lambda_{\max
}-\lambda_{\min}}\right\vert }}}, \label{HL}%
\end{align}
which is clearly capable of reaching HL precision $\delta\kappa\sim1/N$,
significantly beyond the SQL. For instance, assuming an ensemble of $N$
spin-$1$ particles, according to the analysis above, the QCRB gives
$\delta\kappa\geq{\frac{1}{\sqrt{\nu}}}{\frac{1}{{TN}}}$, and a proper initial
probe state that saturates this bound is $( {{{\left\vert {j=1,{m}%
=1}\right\rangle }^{\otimes N}}+{{\left\vert {j=1,{m}=0}\right\rangle
}^{\otimes N}}}) /\sqrt{2}$, which is nothing but the N-particle GHZ state, or
the NOON state \cite{boto2000,lee2002quantum}, a cat state of spin-1 particles
all in spin state $m=0$ or $1$.

The HL estimation of $\kappa$ in Eq. (\ref{HL}) with the NOON state discussed
above also requires measurement of parity, i.e., $\hat{P}={\left(
{-1}\right)  ^{a_{0}^{\dag}{a_{0}}}}$, where $a_{0}$ ($a_{0}^{\dag}$) denotes
the annihilation (creation) of a condensed particle in $m=0$ spin component.
If a $\pi/2$ rotation between $m=0$ and $m=1$ states is applied to all
particles at once, we find ${\langle{\hat{P}}\rangle_{\mathrm{NOON}}}={\left(
{-1}\right)  ^{N/2}}\cos\left(  {N\kappa t}\right)  $ \cite{Holland,gm10},
whose unambiguous determination calls for single atom resolution in detecting
$N$. In addition, as with the paired DFS entangled state of Eq. (\ref{Npairs}%
), NOON state with a meaningfully large $N$ remains to be generated. Were it
to become available, such a Schrodinger cat state is known to be extremely
fragile to the environment perturbations or imperfections of the control
protocols \cite{Pellizzari97,escher2011general,dorner2009optimal}.

Recent years have witnessed tremendous progresses in the generations of other
forms of entangled ensembles with increasingly larger $N$
\cite{hosten2016measurement,bohnet2016quantum,luo2017deterministic}. Compared
to the state Eq. (\ref{Npairs}) suggested earlier, an analogous superposition
based on the two particle DFS is
\begin{align}
{\frac{1}{\sqrt{2}}}  &  \left(  \left\vert {m}=1\right\rangle ^{\otimes
{\frac{N}{2}}} \left\vert {m}=-1\right\rangle ^{\otimes{\frac{N}{2}}%
}+\left\vert {{m}=0}\right\rangle ^{\otimes N}\right)  , \hskip 12pt
\label{tf0s}%
\end{align}
which is also immune to stray magnetic fields. The above state reminds us of a
superposition of two twin-Fock states Eq. (\ref{Npairs}), except for a subtle
difference which might make it realizable in a spin-1 atomic condensate
\cite{luo2017deterministic}. The two components of the superposition in Eq.
(\ref{tf0s}), one twin-Fock state and one Fock state, can both be realized
with quality, as the normal Fock state is simply the polarized state of all
atoms in spin $(m=0)$ component.

Yet a more promising many particle entangled state for estimating $\kappa$ is
the balanced atomic spin-1 Dicke state, which is readily available for
experimental applications \cite{zwl18}. Spin squeezed
\cite{hosten2016measurement,cox2016deterministic} will also work along the
similar line of thoughts discussed below. Although spin-1/2 particles cannot
be used because $(j_{z}^{(i)})^{2}=1/4$ leading to a constant generator
$\mathcal{H}=N/4$, different $m$-states of higher spins (pseudo-spins) can be
employed instead.

Dicke states are broadly defined as the eigenstates of collective spin or
angular momentum $J^{2}$ ($\vec J=\sum_{i} \vec j^{(i)}$) and $J_{z}$. For a
spin-1 atomic condensate and assuming all spin components with the same
spatial mode function, the balanced Dicke state with zero magnetization
($J_{z}=0$) is described by the following wave function
\begin{align}
\label{Dicke}\left|  D \right\rangle = \sum\limits_{k = 0}^{N/2} {{2^{\left(
{N - 2k} \right)  /2}}\sqrt{C_{N}^{k}C_{N - k}^{k}/C_{2N}^{N}} \left|  {k,N -
2k,k} \right\rangle }, \,\,\,
\end{align}
in the Fock state basis $|k,N - 2k,k\rangle$ of $k$ atoms each in $j_{z}%
^{(i)}=m=\pm1$ and the other $N-2k$ atoms in $j_{z}^{(i)}=m=0$ spin component
\cite{zwl18}. In the notation of Eq. (\ref{Npairs}), the above basis state
takes the form $|1,1\rangle^{\otimes k}|1,0\rangle^{\otimes(N-2k)}%
|1,-1\rangle^{\otimes k}$ in terms of spin-1 eigen-state $|f=1,m\rangle$.
$C_{N}^{k}$ denotes the combinatorial factor of choosing $k$ out of $N$. The
balanced Dicke state possesses exceptional coherence \cite{zwl18}. Its basis
states also forms a DFS, and is thus immune to linear Zeeman shifts from stray
magnetic fields like states Eqs. (\ref{Npairs}) and (\ref{tf0s}).

\begin{figure}[pth]
\includegraphics[width=1\columnwidth]{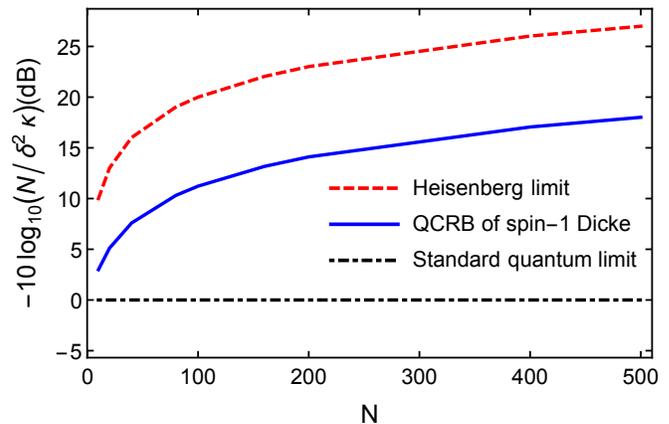} \caption{The quantum
Cram\'{e}r-Rao bound (QCRB) for the balanced spin-1 Dicke state. The black
dot-dashed line denotes SQL ($1/\sqrt{N}$), while the red dashed line refers
to the HL ($1/N$), saturable by NOON state with parity measurement. The blue
solid line shows the QCRB of the balanced spin-1 Dicke state, which clearly
implicates its ability for estimating $\kappa$ beyond the SQL. }%
\label{fig1}%
\end{figure}

\begin{figure}[pth]
\includegraphics[width=1\columnwidth]{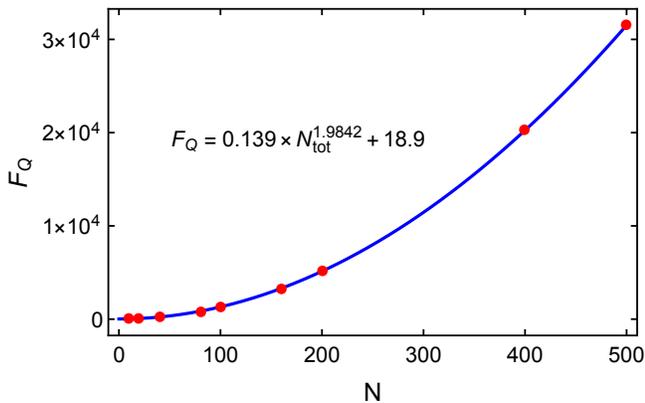} \caption{A numerical fit to
the QFI $\propto N^{\gamma}$ for the estimation of $\kappa$ by the balanced
spin-1 Dicke state gives $\gamma\approx1.98$, implicating that precision at
the HL is possible. }%
\label{fig2}%
\end{figure}

The allowed precision limit for estimating $\kappa$ by the balanced Dicke
state is shown in Fig.~\ref{fig1}, where potential improvement over the SQL,
or $- 10{\log_{10}}({N/{\delta^{2}}\kappa})$, is plotted in blue solid line.
Although HL (red dashed line) is not saturated, its QFI scaling shown in Fig.
\ref{fig2} clearly supports an enhanced precision over the SQL, provided a
suitable measurement scheme is found. Unfortunately, despite of our earnest
effort, we have not been able to find an observable with its associate
measurement protocol to saturate the QFI. If we follow the standard Ramsey
interferometry and measure $J_{x}^{2}$ (or $J_{z}^{2}$ after a $\pi/2$
rotation), we find a precision scaled as $\sim1.48/N^{0.46}$, which
asymptotically approaches the SQL, but not beyond. Nevertheless, we consider
this an encouraging result as single atom resolution in number counting is not
required here. Furthermore, the structure of a DFS facilitates prolonged
coherence time.

In more detail, the numerically computed QFI for estimating $\kappa$ by the
balanced spin-1 Dicke state Eq. (\ref{Dicke}) is shown in Fig. \ref{fig2}. The
fitted scaling exponent gives a $\gamma=1.94$ for $F_{Q}\propto N^{\gamma}$ at
large $N$, which indicates that the ultimate precision for its estimation of
$\kappa$ would asymptotically approach the HL.

Our proposal is equipped with at least two advantages: first it can be more
readily applied to systems of large particle numbers; and second the suggested
Dicke states are already available experimentally with more than 10000 atoms
\cite{luo2017deterministic,zwl18}. According to results of Fig. \ref{fig1}, at
$N=10^{4}$ atoms, the optimal improved measurement sensitivity for the LSV
parameter $\kappa$ can be close to 2 (for SQL) to 4 (for HL) orders of
magnitude smaller than experiments with two ions \cite{dfh16,sos18}. The
precision for the LSV parameter $C_{0}^{(2)}$ likewise can be improved
potentially by the same orders of magnitude, according to the proportional
relationship between the parameters $\kappa$ and $C_{0}^{(2)}$,
\begin{equation}
\frac{\kappa}{2\pi}=\frac{\left[  \Delta E/\left(  hC_{0}^{(2)}\right)
\right]  }{\Delta\left(  j_{z}^{2}\right)  }C_{0}^{(2)},
\end{equation}
which is obtained from Eq. (\ref{wet}) with $\Delta E$ $\left(  \Delta\left(
j_{z}^{2}\right)  \right)  $ denoting the energy (angular momentum) deviation
(fluctuation) for the experimentally selected states and $h$ is the Planck
constant. For the balanced Dicke state we suggest with $^{87}$Rb atoms, a
rough estimate using the method of Ref. \cite{sos18,vvf16} gives ${\Delta
E}/{hC_{0}^{(2)}}=8.6\times10^{15}$ Hz and $\Delta\left(  j_{z}^{2}\right)
=1$ for the $m=1$ and $m=0$ states. Thus, assuming a year-long measurement
with $10^{4}$ atoms, $\Delta\kappa\sim10^{-9}$ can be expected in the SQL.
This implies the parameter $C_{0}^{(2)}$ is bound at the level of $10^{-25}$,
which is about two order of magnitude higher than the results reported
previously \cite{dfh16,sos18}.

\section{conclusion}

This paper discusses LSV effect for the directional dependent interaction
(from the SME) of a single bound valence electron. A recent proposal for
estimating its strength employing Ramsey interferometry is described, and
possibilities of reaching enhanced estimation precision beyond the SQL in
terms of entangled particles are studied. Two alternative but less complicated
many particle entangled states are suggested, both capable of reaching the
precision of HL. Of particular relevance, we show that the balanced spin-1
Dicke state, which can be deterministically produced in spin-1 atomic
Bose-Einstein condensate of $^{87}$Rb atoms \cite{zwl18}, lives in a DFS, and
can reach the precision of SQL by collective Ramsey interferometry without
single atom counting resolution. Although its QFI implicates its ability for
asymptotically reaching the HL precision, we have not been able to nail down
an actual observable or measurement scheme for achieving this.

Finally, we note that there exist other LSV terms of the SME that
reduce to interactions
proportional to the z-component of Pauli operators \cite{kl99},
 and thus are also testable using our proposed states.
The details of how their associated energy
level shifts arise may be different from our discussed Hamitonian (\ref{Hes}),
and deserve further investigations.

This work is supported by NSFC (No. 91636213 and No. 11654001). L.Y. also
acknowledges the support of the National Key R\&D Program of China (Grant No.
2018YFA0306504) and the NSFC (Grant No. 11747605).

\end{document}